\documentclass[preprint,11pt]{elsarticle}
\usepackage{lineno,hyperref}
\usepackage{natbib}
\usepackage{amssymb,amsmath}
\usepackage{algorithm,algpseudocode,algorithmicx,tabularx}
\usepackage{soul}
\usepackage{xcolor}
\usepackage[margin=1in]{geometry}
\usepackage{savesym}
\usepackage{amsmath}
\usepackage{url}

\usepackage{hyperref}
\hypersetup{breaklinks=true}
\usepackage{subcaption}
\savesymbol{iint}
\usepackage{txfonts}
\usepackage{multirow}
\usepackage{tikz}
\restoresymbol{TXF}{iint}
\usepackage{wrapfig}
\usepackage{graphicx}
\usepackage{url}
\usepackage{algorithm}
\usepackage{algpseudocode}
\modulolinenumbers[5]
\journal{Arxiv }

\begin{document}

\begin{frontmatter}

\title{Web3.0 Security: Privacy Enhancing and Anonym Auditing in Blockchain-based Structures
}

\author[rvt]{Danyal Namakshenas}
\ead{dnamaksh@uoguelph.ca}

\address[rvt]{ 
Master of Cybersecurity and Threat Intelligence,\\
University of Guelph, ON, Canada
} 


\begin{abstract}
The advent of Web 3.0, underpinned by blockchain technologies, promises to transform the internet's landscape by empowering individuals with decentralized control over their data. However, this evolution brings unique security challenges that need to be addressed. This paper explores these complexities, focusing on enhancing privacy and anonymous auditing within blockchain structures. We present the architecture of Web 3.0 based on the blockchain, providing a clear perspective on its workflow and security mechanisms. A security protocol for Web 3.0 systems, employing privacy-preserving techniques and anonymous auditing during runtime, is proposed. Key components of our solution include the integration of privacy-enhancing techniques and the utilization of Tor for anonymous auditing. We discuss related work and propose a framework that meets these new security requirements. Lastly, we offer an evaluation and comparison of our model to existing methods. This research contributes towards the foundational understanding of Web 3.0's secure structure and offers a pathway towards secure and privacy-preserving digital interactions in this novel internet landscape.
\end{abstract}

\end{frontmatter}

\section{Introduction}
One of the recent new and hot areas for research and investigation in our digital world is related to Web 3.0 \cite{a1}. With the advent of Web 3.0, which encompasses the semantic web and decentralized technologies, cybersecurity has emerged as a critical focus area \cite{a2}. The concept of a decentralized web seeks to radically transform our digital interactions, introducing both unique challenges and opportunities for ecosystem security. Web 3.0, a potential future iteration of the internet, leverages public blockchains primarily recognized for enabling cryptocurrency transactions. The allure of Web 3.0 lies in its decentralization, wherein users bypass corporate intermediaries like Google, Apple, or Facebook to access the internet, instead personally owning and managing their internet segments. Essentially, Web 3.0 utilizes a fresh suite of blockchain-based technologies that aid in the creation of decentralized web applications, granting users command over their identity, content, and data.

If we want to have a deeper sight into Web 3.0, blockchain is a pounding heart for Web 3.0  and is a key solution to delivering Web 3.0 services. Indeed, the blockchain protocol is the foundational layer of Web 3.0 since blockchain, by nature, can provide security and privacy in any environment \cite{b1}. Many applications and technology, such as IoT (Internet of Things), SDN (Software-Defined Networking), and NFV (Network Functions Virtualization) are combined with that to tackle security threats \cite{a4,a5}. Also, these days, generally accepted that security is one of the most critical issues, and the security concept will be such a crucial problem due to the broad scope of Web 3.0  applications. Therefore, in  Web 3.0 security, a vulnerability refers to anything a hacker can leverage to exploit the protocols in this area. For example, this might be something in the blockchain structure, a flaw in the underlying code, or utilizing blockchain features and characteristics.

Also, in moving beyond  Web 2.0,  Web 3.0  should resolve many of the inherent vulnerabilities in  Web 2.0  technology. However, this process is not completely painless as  Web 3.0   brings with it its own set of vulnerabilities and inherits many of the problems of  Web 2.0 \cite{b2}. However, 
 Web 3.0 promises to provide some unique features like protecting identity and data rights by allowing users to be completely anonymous. This is well seen in cryptocurrencies, where user wallets and transactions, whilst fully visible on the blockchain, are not connected to their identity. Therefore, blockchain technology is one of the foundations of  Web 3.0 and should support  Web 3.0  goals while users might not even notice it. Indeed, people will not care about the underlying infrastructure and its security; they focus on the services and protocols which are used.

Additionally, the security of Web 3.0 can hinge on the vulnerabilities of smart contracts. These are programs or scripts operating on the blockchain, adhering to predetermined rules when certain conditions are met. If a smart contract's code is susceptible to attacks, it can have grave consequences on Web 3.0 services. This issue is further complicated by the absence of legal precedents safeguarding smart contracts. Thus, in many instances, it's impossible to insure or recover potential losses, such as cryptocurrencies and NFTs, in the event of a cyber attack. One of the recent new and hot areas for research and investigation in our digital world is related to Web 3.0. With the advent of Web 3.0, which encompasses the semantic web and decentralized technologies, cybersecurity has emerged as a critical focus area \cite{a6}. As the decentralized web aims to revolutionize how we interact with the digital realm, it brings forth new challenges and opportunities in terms of securing the ecosystem \cite{a7,a8}.

Our motivation at this point is shaped by the advent of Web 3.0, necessitating the exploration of alternative blockchain platforms and protocols that meet emerging security demands. Consider the Oasis Network, which identifies itself as "the first scalable, privacy-centric blockchain." Its Oasis Protocol facilitates 'data tokenization,' a feature that ensures user control over data usage. The organization asserts that this capability will pave the way for more user-friendly Decentralized Finance (DeFi) applications. Findora is another blockchain platform and protocol that merges "transactional privacy" with selective information disclosure to regulators and auditors. This distinguishes it from privacy-oriented cryptocurrencies, or 'privacy coins' like Monero, often the go-to for top-tier ransomware criminals due to its untraceability. Findora secured an "eight-figure" investment round last year, and in October, it introduced a 100 million fund to strengthen its developer community. \cite{b5}.

In other words, motivated by these items, we move forward that should design a security protocol for  Web 3.0 during getting services from blockchain-based structures. To be secure and remove privacy issues, in  Web 3.0 scope, it is necessary to utilize security protocols that provide more privacy and anonymity, especially in blockchain environments \cite{b3}. An auditing feature in  Web 3.0 scope an excellent tool to assess a user's operations and ensure that the records are as accurate as possible. While information sourced externally and internally should have a privacy-preserving feature \cite{a9}. As a result, our goal is secure the entire  Web 3.0 architecture during getting services from blockchain layers that proposed protocol provide professional privacy, security, and audit \cite{a14}.

Despite the considerable attention Web 3.0 has garnered, precise definitions and design outlines are still largely lacking. Some studies have examined consensus-level aspects, but they haven't given a comprehensive view of other equally vital components and architectural designs intrinsic to Web 3.0. This lack of clear definitions and consensus suggests that Web 3.0 is either a concept overhyped without practical development or has multiple potential directions for growth. This paper bypasses broad discourse, instead focusing on the architecture of Web 3.0 and its interplay with blockchain. To summarize, the paper makes the following contributions:

\begin{itemize}
    \item Design and present Web 3.0 architecture based on blockchain structure. The workflow and security mechanisms of Web 3.0 has been clear.
    \item	Proposed a security protocol for Web 3.0 system that utilizes privacy-preserving and anonym auditing in the run-time.
    \item	Applying privacy enhancing techniques as
    \item	Using Tor for anonym auditing
\end{itemize}

Following is a breakdown of the rest of the paper. In Section 2, we discuss work related to blockchain and Web 3.0. Section 3 describes the Web 3.0 System Architecture. In Section 4, the Anonym auditing in Web 3.0, and in Section 5, Evolution of Web 3.0. Section 6 presents the integration of Web 3.0 with New Technology, and conclusions are drawn, and research is suggested going forward.

\section{Related work}\label{ss}
Web3 has become a dominant concept from 2020 onwards, significantly stimulating the growth of the Internet of Value and Metaverse. However, there still lacks robust protocols, security methodologies, and standards in this field. This section will scrutinize the most pertinent research concerning Web3 protocols and security mechanisms in blockchain-based environments. The study in \cite{w3} introduces a Web3 protocol deemed secure if a user can retrieve the correct state and transaction on the blockchain anytime post-block confirmation. The security model is predicated on a strong blockchain, with security assured by persistence and liveness. Persistence refers to uniform views among different nodes at a specific block height, while liveness focuses on the finality of a block within the valid longest chain. We disregard other chain structures like the directed acyclic graph (DAG) \cite{D}.

In another research, the scholars in \cite{w2} illustrate a potential foundation for Web3, elucidating its core components, design principles, and the expansive Web3 design space. They propose that any Web3 execution can be explicated through three primary elements: tree, ledger, and cloud, which adhere to a set of Web3 architectural principles, hence creating a comprehensive Web3 design space. Additionally, \cite{w4} describes a Security Protocol for distributed IoT Microservice, where they apply Web 3.0 technologies in an IoT setting and address its security vulnerabilities using robust security design practices. As explained in Wickström et al.'s research \cite{e}, smart contracts are executed on the Ethereum Virtual Machine (EVM) where user and device authentication and authorization take place. Given the immutable nature of smart contracts, it allows for the creation of a permanent activity log for the protocol. They ensure the privacy of network devices by avoiding the storage of geographical or physical device information in its contracts. In his work, Yang \cite{e2} discusses Timed-Release Encryption in Web3 and provides an Efficient Dual-Purpose Proof-of-Work Consensus, encrypted via an asymmetric key encryption scheme on a blockchain. In a related work \cite{e5}, Marcus contemplates the influence of Web3 on Privacy and Personal Data Management, conducting a thorough analysis of various Decentralized Cloud Storage (DCS) solutions and a case study of interactions with a basic social media application in a Web3 context.

In their study, the authors in \cite{p} propose a distributed protocol for data aggregation enhancing privacy, founded on blockchain and homomorphic encryption. On the other hand, Uzair et al. \cite{r1} put forward a Scalable Protocol for Managing Trust in the Internet of Vehicles with Blockchain, employing smart contracts, physically unclonable functions (PUFs), certificates, and a dynamic Proof-of-Work (dPoW) consensus algorithm. In another investigation \cite{r2}, the authors delve into Smart Contract Deployment on the Ethereum Platform using Web3.js and Solidity with Blockchain, concluding with a methodological approach for the creation, deployment, and interaction of smart contracts using Node.js, the Web3 library, and Infura API.

\section{Web 3.0 System Architecture}


The Web 3.0 system architecture is designed to be open, allowing developers to build applications that are compatible with various devices and systems \cite{a12,a13}. It also integrates various services. Additionally, Web 3.0 System Architecture supports the development of user-generated content and the seamless integration of social networks and other communication protocols. This allows for developing applications that can be accessed from various devices and systems. Based on Web 3.0  concepts and its security goals, we design an architecture for the Web 3.0 system, as presented in Fig .1. This system architecture shares between users and blockchain infrastructure  with the Web 3.0 protocol to deliver Web 3.0 services. In this regard, Fig 1 also illustrates the workflow of the Web 3.0 system.The processing and storage of user data in Web 3.0 is executed in a decentralized and community-controlled network via open protocols, moving away from a centralized TTP \cite{a15}. An enticing feature of Web 3.0 is its system of immediate rewards, ensuring users receive an equitable proportion of revenue when they contribute to the network.

The proposed system structure is logically segmented into four distinct layers: blockchain, application, client, and wallet. The blockchain layer provides enduring ledger storage via mainnet nodes, and its smart contracts deliver the capability to access and modify on-chain data. Distributed storage solutions such as IPFS and Swarm incentivize clients with blockchain tokens for providing distributed storage services. This layer comprises single, homogeneous, and heterogeneous blockchains.

The application layer implements the necessary business and interaction logic for the client layer and centrally stores data that does not require on-chain documentation. This layer encompasses various Web3 systems and employs adaptively scalable technology to "upgrade" the underlying blockchain of Web3, thereby augmenting the usability of Generic-NFT \cite{4s}. The client layer presents web browser interfaces for operations such as minting, buying, and selling NFTs. Every client connects to the blockchain using a web wallet like MetaMask and employs it for signing client transactions. Clients are tasked with handling user requests. For a minimal amount of requests, traditional blockchain systems can engage a single browser as the client to interact with the wallet. However, when there's a sharp increase in requests within a short timeframe, an agent is necessitated to manage the sudden surge of requests. A browser-based wallet provides an intuitive method for users to adopt Web3 services. Users only need to add an extension tool to their browser and import their private key into this embedded wallet. When visiting a Web3-supported website, users can directly connect the wallet, and any functions clicked on the website will invoke the backend methods through APIs under the user's account. An agent-based wallet, on the other hand, facilitates batch processing during high-density user request scenarios. Similar to traditional Web1/Web2, users should initially grant a trusted agent with appropriate permissions. The authentication procedure is initiated once users formally register with the agents.

Protocols are essential to the security of Web 3.0. They act as a set of rules governing how data is transferred between networks, ensuring that data remains secure and confidential. Protocols provide a secure connection between the client and the server when sensitive data is being transmitted. Protocols also provide authentication, which is a way of verifying the user's identity. Authentication requires users to enter a unique username and password that is only known to them. This prevents unauthorized access to data and protects against malicious attacks. Additionally, protocols allow for data encryption, ensuring that the data is unreadable or unusable by anyone other than the intended recipient. This prevents attackers from intercepting data, as they won’t be able to decipher the encrypted information.
Without protocols, the web would be much less secure, as attackers could easily access and use sensitive information.
In the following, we clarify the entities participating in Web 3.0 security protocols.

 \begin{figure}[H]
    \centering
    \includegraphics[width=0.67\linewidth,keepaspectratio]{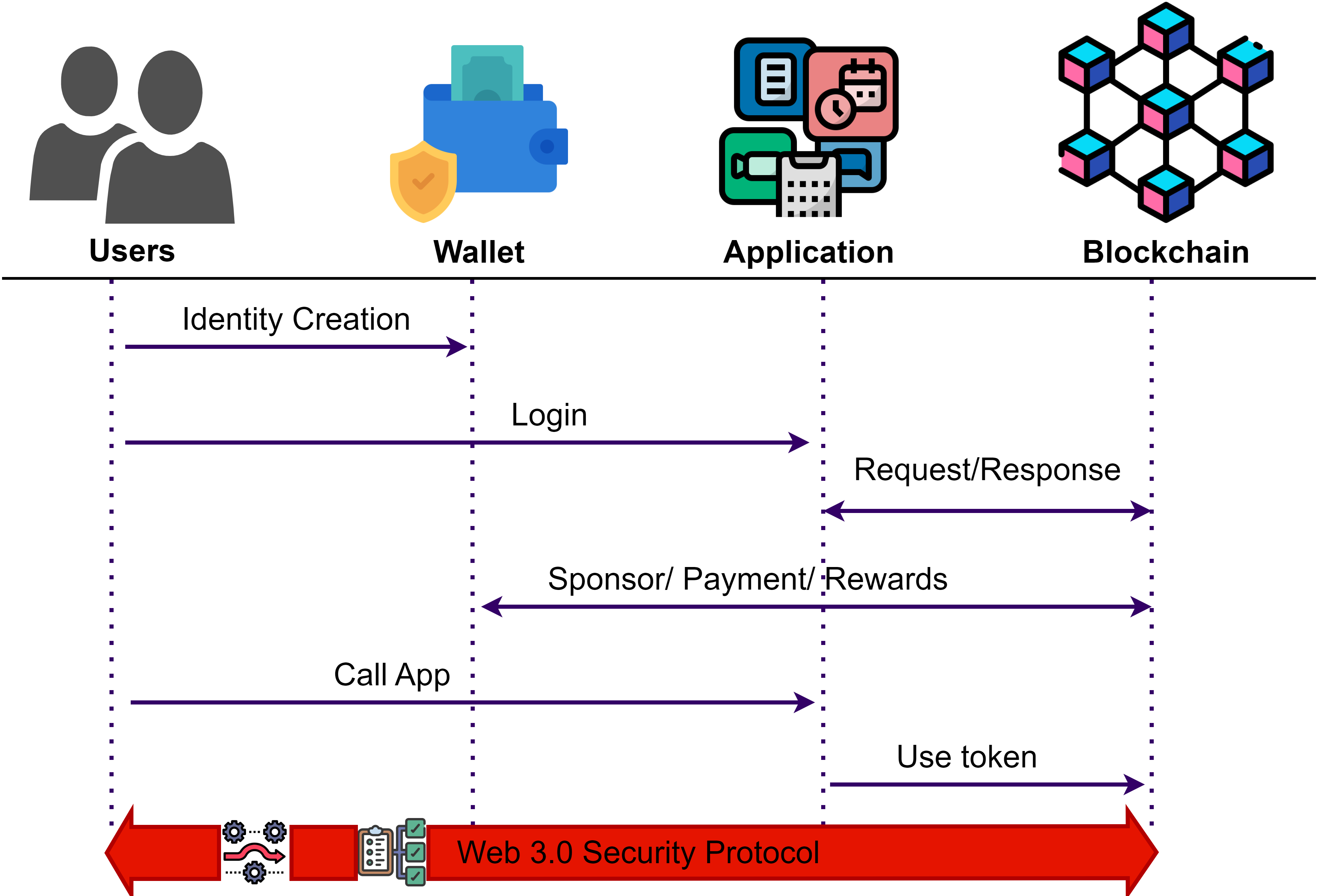}
    \caption{Workflow of A Web3 System}
    \label{fig11}
\end{figure}

\subsection{Privacy Enhancing}

Blockchain transactions empower users to manage their data via private and public keys, thereby giving them ownership. This technology prevents third-party intermediaries from misusing or gaining unauthorized access to the data. When personal information is stored on the blockchain, the owners can dictate the conditions under which third parties can access it. Moreover, blockchain ledgers intrinsically come with an audit trail that guarantees the validity of transactions. Web 3.0 deploys privacy-enhancing technologies with the intent to safeguard user data and privacy while simultaneously facilitating secure and personalized online interactions.One method for achieving this is through the use of decentralized identity systems. Decentralized identity systems use blockchain technology to enable individuals and organizations to own and control their own digital identity. This is in contrast to traditional centralized identity systems, in which a single entity, such as a government or corporation, holds and controls the data.

One example of a decentralized identity system is the Identity Hub. The Identity Hub allows users to store their personal information, such as name, address, and date of birth, on a decentralized platform. This information is then encrypted and stored on the blockchain, making it secure and immutable. Users can then choose to share their personal information with third parties, such as websites or online services, through the use of identity credentials. These credentials, which can be in the form of a digital certificate or a blockchain-based token, allow users to prove their identity without revealing all of their personal information. Decentralized identity systems also offer the ability to manage and revoke access to personal information. If a user no longer wishes to share their information with a particular party, they can simply revoke the credentials, effectively cutting off access to their data.
Another privacy enhancing technology in Web 3.0 is the use of zero-knowledge proofs. Zero-knowledge proofs allow one party, known as the prover, to prove to another party, known as the verifier, that they possess certain knowledge or information without revealing the actual knowledge or information.
For example, a user may want to prove that they are over the age of 18 to a website without revealing their actual age. Through the use of a zero-knowledge proof, the user can prove that they are over 18 without revealing their actual age.
Zero-knowledge proofs have the potential to greatly enhance privacy in online transactions, as they allow individuals to prove their identity or qualifications without revealing sensitive personal information.

In addition to decentralized identity systems and zero-knowledge proofs, Web 3.0 also introduces the concept of self-sovereign identity. Self-sovereign identity refers to the idea that individuals should have full control and ownership over their own digital identity and personal data. This concept is based on the principle that individuals should have the right to determine how their personal information is collected, used, and shared. Self-sovereign identity systems allow individuals to store their personal information on a decentralized platform and control access to it through the use of identity credentials and smart contracts. Overall, privacy enhancing technologies in Web 3.0, such as decentralized identity systems, zero-knowledge proofs, and self-sovereign identity, offer the potential to greatly enhance privacy and security online while still allowing for personalized and secure experiences.

\subsection{Formulate Privacy}

Privacy is a fundamental and intricate concept, quintessential for maintaining the balance of personal freedom and societal interplay in our information-rich society.
Privacy can be formulated as the right of an individual or group to seclude information about themselves, and thereby express themselves selectively. Its importance spans from the social dimension to personal, ethical, political, and legal facets. In formal terms, privacy is multidimensional, embracing various aspects like information privacy, physical privacy, and decisional privacy. Information privacy refers to the rights an individual has to control the collection and usage of their personal information. Personal information encompasses both Personally Identifiable Information (PII) and non-PII, which include any information that can be used directly or indirectly to identify an individual. Physical privacy, on the other hand, pertains to an individual's right to maintain their physical space and personal property free from intrusion. It covers the right not to be subjected to unlawful searches or seizures and the right to bodily integrity.

Decisional privacy refers to the individual's right to make personal choices without governmental intervention. This includes making decisions about one's own body, family, and lifestyle without outside interference. Within the sphere of a data-centric world, privacy often gravitates around principles such as data minimization, purpose limitation, storage limitation, accuracy, integrity, and confidentiality.
Data minimization stresses the collection of only the minimal personal data that's necessary for a particular objective. Purpose limitation necessitates that personal data is gathered for well-defined, explicit, and legitimate purposes, and not processed in a manner incompatible with those purposes. The principle of storage limitation suggests personal data should not be retained in an identifiable form for longer than needed for the purposes for which the data is processed.
The principle of accuracy underscores the importance of maintaining personal data that is both accurate and up-to-date, while the principle of integrity and confidentiality ensures that personal data is handled in a manner that provides sufficient security. This includes protection against unauthorized or illegal processing, as well as against accidental loss, destruction, or damage.

\subsection{Privacy Preserving}

Privacy preserving is a critical concern in the digital age, which deals with the protection of individual or collective information in an environment of proliferating digital data. In a formal sense, privacy preserving techniques are methodologies designed to protect personal or sensitive data from unauthorized access or disclosure, while still allowing useful computations on the data. This concept is instrumental in fields like data mining, machine learning, and cloud computing, where vast amounts of data are analyzed to glean insights while still maintaining the privacy of the individuals represented in the data.

Differential privacy stands out as a significant technique for preserving privacy. It provides a mathematically-backed method to measure privacy leakage. By adding noise to data query results, it ensures that an individual's data presence or absence doesn't greatly influence the output, thereby providing robust privacy assurances.
Other techniques for preserving privacy encompass homomorphic encryption and secure multi-party computation. Homomorphic encryption permits operations to be executed on encrypted data without the need for decryption. Secure multi-party computation, on the other hand, enables multiple parties to perform computations on their combined data, all the while keeping their individual inputs hidden from each other.
Furthermore, anonymization and pseudonymization are key strategies in privacy preservation. Anonymization involves the complete removal of personally identifiable information from data sets, making the identification of individuals impossible. Pseudonymization, however, replaces identifiers with pseudonyms, permitting the identification of individuals under certain conditions.

\section{Anonym Auditing in Web 3.0}
Anonymous auditing is a critical element of ensuring privacy compliance and confidentiality in a data-driven society. It pertains to the inspection and evaluation of systems, processes, or data sets to ensure that privacy laws and standards are being met, while the identity of the individuals remains concealed. Auditing anonymously incorporates techniques that preserve privacy to ensure that the identity of data subjects remains undisclosed during the audit. Techniques such as k-anonymity, l-diversity, and t-closeness are utilized to keep the anonymity of individuals in the dataset intact, yet provide valuable information for the audit. K-anonymity safeguards each individual in a released dataset by making them indistinguishable from at least k-1 other individuals. L-diversity takes it a notch higher by ensuring that within each group of k-indistinguishable individuals, there are at least 'l' diverse and "well-represented" sensitive attributes. T-closeness furthers the concept by requiring that the distribution of a sensitive attribute within any equivalence class closely mirrors the attribute's distribution in the complete dataset.

\subsection{Formulate Auditing}

Defining auditing within a formal framework requires a clear understanding of its fundamental purpose and methodologies. Essentially, auditing is a standalone, objective assurance and consulting activity aimed at enhancing and refining an organization's operations. By employing a systematic, disciplined approach, it aids an organization in achieving its goals through the evaluation and enhancement of risk management, control, and governance processes.
When applied to information systems, auditing is founded on several essential principles. The first is the principle of independence, which asserts that the auditor must maintain an objective distance from the process, system, or organization under audit to guarantee an unbiased assessment. The second principle is evidence-based reporting. This principle obliges auditors to root their findings and conclusions in the evidence collected throughout the audit process. This ensures the reliability of the conclusions and their ability to withstand critical examination.

The third is the principle of relevance. The audit must focus on those aspects that are relevant to the audit scope and objectives, and have potential to improve the organization's operations. The formulation of auditing also encompasses the use of various audit methodologies and tools. These may include control self-assessment, risk assessment, benchmarking, and data analysis. In the digital age, auditors also use advanced technologies such as artificial intelligence, machine learning, and data analytics to analyze large volumes of data and detect anomalies or trends.
In the context of data privacy, auditing is formulated around privacy laws and regulations such as the General Data Protection Regulation (GDPR) in Europe, the California Consumer Privacy Act (CCPA) in the United States, and others. Auditors evaluate the organization's compliance with these laws and regulations, and assess the effectiveness of the organization's data protection controls. This may include evaluating the organization's data privacy policies, procedures, and practices, as well as the technical and organizational measures in place to protect personal data.

\subsection{Auditing in Web 3.0}
The inherent transparency and immutability of blockchain technology make it a viable tool for auditing and verifying transactions in Web 3.0 systems. The term "audit" traditionally refers to an official examination of an organization's accounts or systems, typically by an independent body. However, in the context of Web 3.0, the concept of auditing extends beyond just financial transactions. Auditing in Web 3.0 systems can include the review and verification of smart contract execution, compliance with decentralized governance protocols, and the confirmation of data integrity within decentralized applications (dApps).
One of the critical benefits of auditing in Web 3.0 is the potential for "real-time auditing," where transactions and activities on the blockchain can be monitored and verified continuously. This is facilitated by the public availability of all transactions on the blockchain, which allows auditors, regulators, and users to view and confirm transactions as they are added to the blockchain.
Furthermore, with the integration of zero-knowledge proofs and other cryptographic tools, auditing in Web 3.0 can provide assurance of the integrity and authenticity of transactions without compromising the privacy of the parties involved. This presents a significant advancement over traditional auditing methods, which often involve intrusive inspections and potential privacy risks.
In addition to these benefits, Web 3.0 auditing could also support decentralized governance models. Many blockchain platforms and Web 3.0 applications are implementing decentralized governance protocols, where token holders or network participants can vote on proposals and decisions regarding the network. Auditing plays a crucial role in these governance models by verifying the legitimacy of votes and enforcing compliance with the agreed-upon protocols. To leverage the benefits of Web 3.0 auditing, various tools and frameworks have been developed. For instance, blockchain explorers allow users to view and track transactions on the blockchain, while smart contract analysis tools can help in the detection of vulnerabilities and the verification of smart contract behavior. Also, decentralized auditing platforms are emerging, which aim to provide independent auditing services for blockchain-based systems and dApps.

However, despite its potential benefits, auditing in Web 3.0 also presents several challenges. For one, the complexity and technical nature of blockchain technology and smart contracts can make the auditing process difficult for those without specialized knowledge. Moreover, while transparency is a key feature of blockchain technology, it also raises privacy concerns that need to be addressed. The integration of privacy-preserving technologies, like zero-knowledge proofs, could help alleviate these concerns, but these technologies are still in their early stages and may introduce additional complexity. As Web 3.0 systems continue to evolve, the role of auditing will become increasingly significant. Not only can auditing provide assurances of integrity and compliance in these systems, but it could also play a vital role in supporting decentralized governance models and facilitating the wider adoption of Web 3.0 technologies.

\section{Evolution of Web 3.0}
The onset of Web 3.0 has engendered transformative digital experiences, facilitating semantic data interoperability and collaborative networks. In conjunction with these advancements, Web 3.0's security dynamics are evolving to accommodate new paradigms of privacy and anonymity, particularly in blockchain-based structures. This evolution encompasses privacy-enhancing techniques and anonymized auditing measures that fortify the confidentiality and integrity of information exchanges.
Privacy in Web 3.0 is no longer a mere supplement; it is a foundational necessity. The decentralization inherent to the blockchain technology imbues the Web 3.0 with a fundamental shift from centralized data repositories to peer-to-peer networks. This scenario necessitates robust privacy-enhancing techniques that protect users' sensitive information without jeopardizing the collaborative essence of Web 3.0. Differential privacy, homomorphic encryption, and secure multi-party computation have emerged as paramount techniques that enable computations on encrypted data, mitigate privacy leakage, and allow collaboration while preserving privacy.

Differential privacy introduces algorithmic noise into data queries to guarantee that the output is minimally affected by the inclusion or exclusion of an individual's data. Homomorphic encryption allows data to remain encrypted while computations are performed, obviating the need for decryption and thereby reducing the risk of privacy breaches. Secure multi-party computation enables multiple parties to collaborate on data analysis without revealing individual inputs, fostering a shared information environment that upholds individual privacy. Concurrently, the ascendance of blockchain technology in Web 3.0 ushers in an era of transparent, immutable transactions. While this transparency fosters trust and accountability, it could potentially infringe on users' privacy if their identities are associated with their blockchain transactions. Anonymity becomes a vital counterweight to maintain privacy, prompting the development of anonymized auditing methodologies.

Anonymous auditing in blockchain-based structures involves inspecting and validating transaction integrity and compliance with privacy standards while keeping the identities of the parties involved concealed. Techniques such as k-anonymity, l-diversity, and t-closeness are implemented to provide statistical guarantees of privacy. These methodologies ensure that sensitive attributes within datasets are well-represented and that the distribution of a sensitive attribute in any equivalence class closely approximates the overall dataset distribution.

In essence, the evolution of Web 3.0 security is characterized by a dynamic interplay between privacy enhancement and anonymous auditing within the blockchain ecosystem. The objective is to construct a secure, collaborative, and privacy-preserving environment where users can engage with digital services confidently. As we continue to traverse the terrain of Web 3.0, these security measures will become increasingly sophisticated and integral to the architecture of the decentralized web.

\section{ Integration of Web3.0 with New Technology} \label{tech_integration}
Web 3.0, also known as the semantic or decentralized web, is designed to be more secure, private, and efficient than its predecessor. It incorporates cutting-edge technologies like blockchain, machine learning, and Internet of Things (IoT) devices to create a more streamlined and personalized online experience. Among the key technologies poised to play a significant role in the development and operation of Web 3.0 are Software-Defined Networking (SDN), Federated Learning (FL), and IoT.

\subsection{Software-Defined Networking (SDN)}
The core principle of SDN lies in its separation of the control plane from the data plane. This separation allows network administrators to shape traffic from a centralized control console without having to touch individual switches within the network \cite{a16}. The control plane, responsible for deciding how packets should be forwarded, communicates with the data plane that carries out these decisions, enabling a more dynamic and responsive network architecture \cite{n1}.
In the context of Web 3.0, this separation becomes crucial. As the number of devices and applications within the network grows, the complexity of managing data flow increases exponentially. SDN, with its centralized control, can manage this complexity effectively, ensuring that data packets are routed optimally, reducing latency, and improving the overall performance of the network \cite{n2}. Moreover, SDN's programmability extends beyond simple traffic management. It can be used to implement sophisticated network functions, such as load balancing, intrusion detection, and firewalling, directly into the network infrastructure \cite{a17}. These functions can be crucial in a Web 3.0 environment, where security and reliability are paramount. By integrating these functions into the network control, SDN can provide a robust and secure infrastructure for the decentralized Web 3.0 applications \cite{n2, a18}.

Furthermore, the abstraction provided by SDN can be beneficial for blockchain networks. By abstracting the network infrastructure, SDN allows blockchain nodes to focus on their core tasks, such as transaction verification, without worrying about the underlying network conditions. This abstraction can lead to more efficient use of resources and improved performance of the blockchain network \cite{a17}. In conclusion, SDN's unique features of control-data plane separation, direct programmability, and infrastructure abstraction make it an ideal choice for managing the complex and dynamic nature of Web 3.0 and blockchain networks. Its integration can lead to optimized data pathways, efficient traffic management, and enhanced network performance, thus facilitating faster transaction verification and increasing overall network throughput \cite{n2, a18}.

\subsection{Federated Learning (FL)}
FL's decentralized nature aligns well with the principles of Web 3.0, where data ownership and control are distributed among users rather than centralized authorities. In traditional machine learning models, data from all sources is collected and processed in a central location, which can lead to potential privacy breaches and misuse of data \cite{n3}. FL, on the other hand, keeps the data on the original device and only shares model updates, significantly reducing the risk of data leakage \cite{a19}. Moreover, FL is not just about privacy. It also offers benefits in terms of efficiency and scalability. By training models on the edge devices where data is generated, FL reduces the need for data transmission, which can be a significant bottleneck in large-scale machine learning applications. This can lead to faster model training and lower communication costs, making FL a more efficient and scalable solution for machine learning in a Web 3.0 environment \cite{n4}.

FL also enables more personalized and accurate models. Since the models are trained on local data, they can capture the unique patterns and characteristics of the data at each node. This can lead to more personalized models that can provide better predictions for each user, enhancing the user experience in Web 3.0 applications \cite{a20}.
Furthermore, the integration of FL with blockchain technology can further enhance the security and privacy of the system. Blockchain can provide a transparent and immutable record of model updates, ensuring that no malicious changes are made to the model. This can increase the trustworthiness of the FL system and encourage more users to participate in the learning process, further improving the performance of the model \cite{a25}. FL is a promising approach for machine learning in a Web 3.0 environment. Its ability to train models on decentralized nodes while preserving privacy, reducing communication costs, and providing personalized predictions makes it an ideal solution for the challenges of Web 3.0. By integrating FL with other Web 3.0 technologies like blockchain and SDN, we can create a secure, efficient, and user-centric Web 3.0 ecosystem \cite{a20}.

\subsection{Internet of Things (IoT)}
IoT, or the Internet of Things, refers to a network of physical devices or "things" embedded with sensors, software, and various technologies aimed at connecting and sharing data with other devices and systems via the Internet \cite{n5, a21}. The fusion of IoT with Web 3.0 and blockchain technologies presents intriguing opportunities \cite{n6, a22}. For example, data generated by IoT devices can be logged on a decentralized ledger, enhancing both transparency and security. Blockchain can serve as a trustworthy and unchangeable platform for IoT devices to securely exchange information in a Web 3.0 setting, thereby guaranteeing the authenticity and integrity of the data.
Moreover, smart contracts on the blockchain can be used to automate IoT operations. For example, a smart fridge could automatically order groceries when they're running low, or an IoT-connected car could self-execute leasing agreements. All these transactions can be recorded on the blockchain for transparency and non-repudiation, making IoT devices smarter and more autonomous.
In conclusion, the integration of SDN, FL, and IoT with Web 3.0 can significantly enhance the efficiency, privacy, and functionality of the Web 3.0 environment. SDN can optimize the network performance, FL can provide privacy-preserving machine learning, and IoT can offer seamless connectivity between various devices, all under the secure and transparent umbrella of blockchain technology.

\subsection{Hardware Security for Web 3.0}

While the integration of various technologies plays a crucial role in Web 3.0, it is essential to address hardware security considerations to ensure the overall integrity and resilience of the decentralized web. Hardware security encompasses measures and techniques designed to protect the physical components and underlying infrastructure that power Web 3.0. Here are some key aspects of hardware security in the context of Web 3.0:\\

1- Trusted Execution Environments (TEEs): TEEs are hardware-based security mechanisms that provide isolated execution environments for sensitive computations \cite{Tee}. They offer secure enclaves where critical operations, such as cryptographic key \cite{a23} management and secure transaction processing, can be performed. TEEs protect against attacks targeting the integrity and confidentiality of data, ensuring that critical operations are shielded from unauthorized access.\\

2- Secure Hardware Components: In Web 3.0, the security of hardware components becomes paramount. This includes ensuring the authenticity and integrity of hardware devices, such as IoT sensors and gateways, by implementing secure boot processes, tamper-resistant designs, and secure element integration \cite{a24,n7}. Hardware components should be resistant to physical attacks, such as tampering, reverse engineering, and side-channel attacks, to safeguard the overall security of the Web 3.0 ecosystem.\\

3- Hardware-based Key Storage: As Web 3.0 heavily relies on cryptographic mechanisms to ensure privacy and secure transactions, the secure storage of cryptographic keys becomes crucial. Hardware Security Modules (HSMs) or specialized secure chips can be utilized to store and manage cryptographic keys securely. These hardware-based solutions provide tamper-resistant environments and protect against key extraction or unauthorized use.\\

4- Supply Chain Security: Ensuring the integrity of the hardware supply chain is critical to prevent the insertion of malicious components or tampering during manufacturing or distribution processes \cite{a26}. Implementing mechanisms to verify the authenticity and integrity of hardware components, such as utilizing trusted suppliers, secure manufacturing practices, and tamper-evident packaging, helps mitigate supply chain-related risks\cite{a25, a27}.\\

5- Hardware-based Attestation: Hardware attestation mechanisms allow for the verification and attestation of the integrity and identity of hardware devices. These mechanisms enable trust establishment and can be utilized to ensure that the participating hardware devices in Web 3.0 networks meet the desired security requirements and are not compromised \ref{ffe}.\\

6- Hardware-based Random Number Generators (RNGs): RNGs are essential for various cryptographic operations in Web 3.0, such as key generation and digital signatures. Hardware-based RNGs provide a source of true randomness, which is more secure than software-based pseudo-random number generators. They are resistant to prediction and manipulation, thereby enhancing the security of cryptographic operations \cite{ffe}.\\

7- Physical Unclonable Functions (PUFs): PUFs are unique features of hardware devices that are inherently random and irreproducible, even by the manufacturer. They can be used to generate device-specific cryptographic keys, providing a high level of security against cloning and reverse engineering attacks. PUFs can be particularly useful in IoT devices, which form a significant part of the Web 3.0 ecosystem \cite{cc}.\\

8- Hardware Security in Edge Computing: With the rise of edge computing in Web 3.0, ensuring the security of edge devices becomes crucial. This includes protecting the data stored and processed at the edge, securing the communication between edge devices, and ensuring the integrity of edge computations. Hardware-based security measures, such as TEEs and secure boot, can provide robust protection for edge devices \cite{Tee,a31}.\\

9- Post-Quantum Cryptography Hardware: With the advent of quantum computing, traditional cryptographic algorithms that rely on the difficulty of factoring large numbers or solving discrete logarithm problems can be broken. Post-quantum cryptography aims to develop new algorithms that can resist quantum attacks. Implementing these algorithms in hardware can provide a higher level of security and performance \cite{gg}.\\

10- Hardware Trojans Detection: Hardware Trojans are malicious alterations to hardware components that can cause undesired effects, such as information leakage or system failure. Detecting these Trojans is challenging due to their stealthy nature. Techniques such as side-channel analysis, functional testing, and hardware attestation can be used to detect and mitigate Hardware Trojans \cite{a30}.\\

Hardware security forms a critical pillar of the overall security framework for Web 3.0. By integrating robust hardware security measures, Web 3.0 can ensure the integrity, confidentiality, and availability of its services, providing a secure and trustworthy environment for users. The combination of software and hardware security measures will provide a comprehensive security solution that can withstand the diverse and evolving threats in the Web 3.0 ecosystem \cite{cc,a30}. By addressing hardware security considerations, Web 3.0 can build a robust foundation that enhances the overall security posture of the decentralized web. Combining software-level security measures with hardware-level protections ensures a holistic approach to safeguarding critical operations, data confidentiality, and the trustworthiness of the infrastructure supporting Web 3.0 applications and services.

\section{Conclusion} \label{r5}
In conclusion, the emergence of Web 3.0, backed by blockchain technology, stands to reshape the digital world by shifting power dynamics and offering more privacy and control to users. However, it also brings with it an array of unique security challenges that require comprehensive solutions. This paper shed light on these issues, explicitly defined Web 3.0 architecture based on the blockchain, and proposed a security protocol designed to enhance privacy and support anonymous auditing. Our work presented the practical design and operation of Web 3.0, with a focus on overcoming its inherent security challenges. The proposed security protocol emphasized privacy preservation and anonymous auditing, vital factors in securing Web 3.0 platforms. We also demonstrated the application of privacy-enhancing techniques and the use of Tor for anonymous auditing in our proposed model. The comparative analysis further validated the efficacy of our approach as compared to existing methods. However, as the landscape of Web 3.0 is still evolving, so too must the security solutions we propose. Future work will focus on refining and expanding this protocol as the field continues to evolve. As we transition from Web 2.0 to 3.0, the importance of security and privacy will only grow, and we believe our research contributes significantly to the continuous dialogue in this space.


\bibliographystyle{elsarticle-num}
\bibliography{References}

\end{document}